\RequirePackage{fix-cm}
\documentclass[smallextended]{svjour3}       
\smartqed  
\usepackage{graphicx}
\usepackage{titlecaps}
\usepackage{hyperref}
\usepackage{wrapfig}
\usepackage{footmisc}
\usepackage[justification=centering]{caption}
\begin{document}

\title{Fighting Disaster Misinformation in Latin America: 
}
\subtitle{The \#19S Mexican Earthquake Case Study}


\author{*Claudia Flores-Saviaga         \and
        Saiph Savage 
}


\institute{*Claudia Flores-Saviaga \at
              West Virginia University, WV, USA.\\
              \email{cif0001@mix.wvu.edu}             \\
           \and Saiph Savage
            \at
             West Virginia University, WV, USA.\\
              \email{saiph.savage@mail.wvu.edu}  
}

\date{Received: date / Accepted: date}

\maketitle

\begin{abstract}
Social media platforms have been extensively used during natural disasters. However, most prior work has lacked focus on studying their usage during disasters in the Global South, where Internet access and social media utilization differs from developing countries. In this paper, we study how social media was used in the aftermath of the 7.1-magnitude earthquake that hit Mexico on September 19 of 2017 (known as the \#19S earthquake). We conduct an analysis of how participants utilized social media platforms in the \#19S aftermath. Our research extends investigations of crisis informatics by: 1) examining how participants used different social media platforms in the aftermath of a natural disaster in a Global South country; 2) uncovering how individuals developed their own processes to verify news reports using an on-the-ground citizen approach; 3) revealing how people developed their own mechanisms to deal with outdated information. For this, we surveyed 356 people. Additionally, we analyze one month of activity from: Facebook (12,606 posts), Twitter (2,909,109 tweets), Slack (28,782 messages), and GitHub (2,602 commits). This work offers a multi-platform view on user behavior to coordinate relief efforts, reduce the spread of misinformation and deal with obsolete information which seems to have been essential to help in the coordination and efficiency of relief efforts. Finally, based on our findings, we make recommendations for technology design to improve the effectiveness of social media use during crisis response efforts and mitigate the spread of misinformation across social media platforms.

\end{abstract}

\section{Introduction}
\label{intro}
Social media platforms have transformed how people communicate after a natural disaster \cite{reuter2018fifteen,dailey2017social,gui2017managing}. The information that people share on social media in the aftermath of a natural disaster can be extremely helpful for emergency response teams to identify urgent needs, plan relief efforts and immediately identify affected areas \cite{zade2018situational}. To comprehend this new phenomenon, researchers have begun to study how people utilize social media to collaborate towards a common cause during crisis events \cite{starbird2011voluntweeters}. For instance, prior work identified that people use social media to collectively understand crisis events as they unfold \cite{avvenuti2018need}, filter and classify the content that is shared on social media \cite{zahra2018understanding,alam2018graph,bica2017visual}, and handle information overload \cite{alam2018twitter}.

Social computing platforms (i.e. Facebook, Twitter) give ordinary people the capacity of gathering information to aid on-the-ground emergency response \cite{norris2017digital}, ease the coordination and communication between victims and rescuers \cite{reuter2013combining,kaufhold2016self}, help people to organize fundraisers \cite{muralidharan2011hope}, or even rescue lost pets \cite{white2014digital}.  Social media has even influenced the way governments engage with citizens in the aftermath of a disaster \cite{chatfield2014sandy}. However, few research has been done from a multi-platform perspective. This perspective is important to comprehend how social media relate to each other in a crisis context \cite{dailey2017social}.

Most prior work has focused primarily on studying natural disasters and social media usage in the US \cite{reuter2018fifteen}. The few papers that do study multi-platform use and crisis response in other regions have generally been done from a qualitative perspective \cite{wong2017social}. Many questions remain surrounding how and why people use multiple social media platforms during a disaster. We especially still have a limited understanding of how social media platforms are used in regions where digital and offline volunteers are needed the most, such as developing countries where response groups have limited resources. 

This paper addresses this gap by studying at scale how the general public used different social media platforms in the aftermath of a natural disaster. 
We focus on the analysis of how social media was used during the 7.1-magnitude earthquake that hit Mexico on September 19 of 2017, leaving 6,000 injured and 370 killed. The event is often referred to by its hashtag: \#19S. Notice that it is challenging to study people's online behavior in the aftermath of a natural disaster because it is difficult to track at scale how specific individuals used multiple social media platforms. To overcome this challenge, we follow an approach similar to \cite{wilson2018assembling}, where we  conduct a multi-level analysis at three levels: at the micro-level in which we zoom-in through a survey study and targeted content analysis to understand in detail which social media platforms participants used, why they decided to use them, and how they operated across them. Using temporal charts of volumes of information over time we transition to the meso-level, unpacking how information was assembled the days after the \#19S earthquake. This level allows us to see how information was being generated. At the macro-level, we zoom-out using network representations and descriptive statistics to obtain a more general picture of how the general public used at scale the different social media platforms in the \#19S aftermath. The macro-level allows us to reveal the structure of the information space that takes in the different social platforms analyzed and uncover the broader patterns of information flow. 

Through our analyses, we found that the \emph{reasons} participants used multiple social computing platforms were for: 1) reading, sharing and verifying news reports; 2) sharing needs; 3) receiving updates from friends and family; 4) donating money; 5) building technology; 6) performing data analysis; 7) connecting strangers; and 8) mobilizing people offline.  We identified that participants developed mechanisms for verifying information across platforms and ensuring the information that flowed on social media was relevant. Our macro-level analysis revealed that the information most reshared on Twitter came from organizations that were sharing and verifying news; while on Facebook it came from ordinary people who created Facebook Groups and pages focused on connecting strangers to help them find their missing pets. 


Our work extends the literature of crisis response in the Global South by: 1) examining how participants used different social media platforms in the aftermath of a natural disaster in a Global South country; 2) uncovering how individuals developed their own processes to verify news reports using an on-the-ground citizen approach; 3) revealing how people were able to develop their own mechanisms to overcome the problem of dealing with outdated information.  We believe our findings can help inform designers who wish to create tools to coordinate collective action across different social media sites, powering tools not only for disaster response but also for a range of other endeavors.

\section{Background on Mexico's September 19th Earthquake}
\subsection{The \#19S Earthquake}
On Tuesday, September 19 of 2017, at 13:14 CDT, the \#19S Earthquake struck south of the city of Puebla de Zaragoza, in the state of Puebla, in the southern part of Mexico. Due to the location of the Epicenter, it is often referred to as the Puebla Earthquake. The estimated magnitude of the earthquake was 7.1 and affected the capital, Mexico City, and several states. The total number of casualties was 370 people and over 6,000 people were injured during the disaster \cite{BYUengin7:online}. The earthquake caused over 40 buildings to collapse in Mexico City \cite{BYUengin7:online}. 
During the aftermath, the Mexican Army and Navy deployed 3,000 troops to Mexico City to perform search and rescue missions along with assisting in the cleaning up efforts. Countries across the world also sent their rescue teams, consisting of workers and rescue dogs to join in the rescue efforts \cite{MexicoEa73:online}. 
 

\subsection{{\#Verificado19S}}

Hours after the earthquake struck, a citizen-led initiative  composed of media outlets, companies, NGOs, and universities got together to organize and corroborate information to help strengthen the humanitarian response and provide verified information about the earthquake to anyone interested. The initiative was dubbed: \emph{Verificado19S}\footnote{\url{https://verificado19s.org/sobre-v19s/}} (Verified19S).  Their work led to the creation of the hashtag \#Verificado19S and the Twitter account \textit{@Verificado19S}, a reference to verified information related to the earthquake. The hashtag immediately took off across social media. \emph{Verificado19S} continued to be used by thousands of people in the earthquake's aftermath. It became the most up-to-date source of information about post-earthquake conditions in Mexico, with 36,000 followers on Twitter \cite{Verifica57:online}, and over 500 volunteers on the ground \cite{Verifica17:online}. The news media started to report how these civic response groups were organizing \cite{Tecnolog99:online}. Several news reports covered how these individuals were organizing on Facebook, Twitter, Whatsapp, and Slack \cite{Abiertoa84:online,Verifica57:online}. We built off of these initial news reports to better understand how people used a variety of social platforms during a natural disaster. 


\section{Related Work}

\subsection{Multi-platform Citizen-Led Crisis Response} Social media platforms are increasingly being used during disaster events. These platforms are frequently used by affected people, emergency responders, and volunteers to share and seek information, and provide numerous forms of support \cite{zade2018situational,wong2017social,olteanu2015expect,chauhan2017providing,bica2017visual,alam2018twitter}. However, how these social media platforms are used jointly during a disaster is still understudied \cite{grace2018community}, especially in the Global South where Internet access and social media usage differs from developing countries \cite{nielsen2018empirical}. Researchers have drawn attention to the importance of understanding and conceptualizing online social media, as an
ecosystem of related elements \cite{dailey2017social}. They highlight that  studying  and viewing these interactions at a variety of resolutions, can enhance our understanding of how people process information and make use of platforms, and, thus, shed light on why particular platforms are used and for what reasons \cite{creswell2017designing}. Approaching social media as a holistic ecosystem will provide us with more realistic analyses \cite{SocialMe25:online}.

Researchers have started to analyze how individuals perform information work across platforms during crisis events and how different actors interact on these platforms. However, most prior work has lacked focus  on studying social media usage in the Global South \cite{reuter2018fifteen}.  Gui et al.  analyzed people's conversations about making travel decisions in response to the Zika virus crisis on three online forums: Reddit, BabyCenter, and TripAdvisor, performing a qualitative study of personal risk assessment on travel-related decision making during the crisis. In their work, researchers stress that their studies  might have missed information from Latin America, where English is not the primary language \cite{gui2017managing}. In a recent example from the US that attempts a broader analysis of this kind, Dailey and Starbird \cite{dailey2017social} combined on-site interviews and trace ethnography to follow information work across multiple platforms to analyze how they were used after the 2014 Oso Landslide in Washington state. In Europe, researchers examined the use of social media during the European Floods of 2013, finding that Twitter was used for status updates, while Facebook gave a situational overview to coordinate virtual and off-line activities \cite{reuter2015xhelp}. Another study in Ecuador investigated how specific individuals used different social computing platforms for crisis response in an earthquake \cite{wong2017social}, however; their study was mainly qualitative with a small number of participants. A recent research work of the earthquake in Mexico studied in this paper only did a temporal analysis of the media coverage using Twitter data \cite{curiel2019temporal}. These works provided rich insights about how and why people used different social computing to support informal, on-the-ground, crisis response; however, they are
still limited in scope.

\subsection{Information Sharing During Natural Disasters} Timely and accurate communication is essential during a crisis event \cite{chauhan2017providing}, as it can facilitate relief and recovery efforts, and reduce anxiety and fears \cite{fearn2016crisis}. A body of research has examined how people seek and share local information online during crises \cite{palen2018social,fearn2016crisis}. Speed of information sharing renders social media particularly susceptible to the spread of misinformation \cite{stephens2019new}, due to the factors of information scarcity and ambiguity during these type of events \cite{palen2018social}.  One reason for the spreading of misinformation during disasters is that it can be challenging for people to understand what information can be trusted amidst all this socially-generated data \cite{chauhan2017providing}. Other researchers have argued that spreading misinformation is part of the collective sense-making process that occurs during the crisis \cite{starbird2016could}. Misinformation during disasters has been a major limitation to the use of social media content in the decision making of emergency respondents \cite{wong2017social}. Past research studies have examined the spread of misinformation on social media during disasters. Starbird et al \cite{starbird2014rumors} studied rumors in the aftermath of the 2013 Boston Marathon Bombings, founding evidence that the online crowd identified and corrected rumors, but they noted that corrections often lagged behind the misinformation. 
In a past study, Chauhan and Hughes \cite{chauhan2017providing} used the 2014 Carlton Complex Wildfire to explore who contributes official information online during a crisis event, finding that local news media played an important role in distributing official crisis information online. Another study \cite{starbird2018engage} found that news media are also more likely to ``author original tweets and to be retweeted'' by others. This means that they play an important function in ``shaping the news.'' However, the ability of news media to keep up with the speed of demand for information in a world dominated by social media means that the dictate ``verify, then publish \cite{kovach1999warp}'' suffers greatly.\\
In this paper, we adopt a multi-level perspective that encompasses different lenses across various platforms investigating the distinctive structural content and temporal aspects of the online interactions of individuals during an earthquake in the Global South. Our analysis provides a deeper examination as to how these platforms are cross-referenced and used. We accomplish this by paying particular attention to the collaborations between participants in order to support an information verification process during the earthquake and how their activities and information flow from one platform to the other.

\vspace{-0.8pc}
\section{Methods}

Acknowledging the scale and multi-sited nature of the networked discourse that we study, we bootstrap off methods for conducting research on online interactions and collaborations in crisis events \cite{palen2016crisis}, network ethnography \cite{arif2018acting}, and collaborative work within online communities \cite{wilson2018assembling} that have had to work with similar contexts. In specific, we conduct a multi-level analysis that allows us to first zoom-in and obtain the details of how participants used different social media platforms during and after the earthquake, and then use those findings to inform a larger scale study that provides a broad picture about how different social media platforms were used for \#19S. Similar to prior work \cite{wilson2018assembling}, the multi-levels we study are: 1) micro; 2) meso; and 3) macro levels.

\subsection{Methods: Micro-Level} Our goal at the micro-level was to zoom in and understand how and why participants of the survey used multiple platforms for \#19S. We performed online surveys to ask people which platforms they used, as well as the reason they had for using them. The survey included questions about the following:

\begin{enumerate}
\item {Demographics and background such as age, gender, and location.}
\item {Open-ended questions about:}
\begin{itemize} 
        \item {How they contributed to the relief efforts.}
        \item {How they used technology to help relief efforts.}
        \item {Their perceptions on the impact of their contributions.}
\end{itemize}
\item {Multiple choice questions about the platforms used, e.g., ``how much did you use Twitter to help in \#19S?''}
\end{enumerate}

{\bf{Recruitment.}} After the earthquake, news about the relief efforts started emerging. The narrative from the news media was that \#19S volunteers included both ordinary and technical people \cite{Mexicoma53:online,OpinionM43:online}. According to news reports, ordinary people were organizing on social media platforms such as Facebook and Twitter to coordinate efforts \cite{Elactivi21:online}. Meanwhile, a group of technologists launched a website called comoayudar.mx (which translates to ``howtohelp.mx''). In there, they invited people with coding skills to collaborate in technological relief initiatives they were developing. They invited people to join a Slack group \cite{Abiertoa84:online,HowtheVe71:online}, specifically the Slack group of  \emph{Codeando M\'{e}xico}\footnote{\tiny \url{http://slack.codeandomexico.org/}} (Spanish for ``Coding for Mexico''), an NGO focused on developing civic  technology in the country\footnote{\tiny \url{https://www.codeandomexico.org/}} \cite{Codeando9:online}. Given this range of platforms, we aimed to recruit both technical and non-technical individuals. For this reason, we posted invitations to our survey on Twitter, Facebook and Slack five months after the earthquake.

\textbf{Twitter}. Some of the authors tweeted invitations to the survey. Combined, the authors have 6,921 followers, many of whom are in Mexico. In the tweets, we used the hashtags\footnote{\tiny \#sismoCDMX, \#sismoCDMX19Sep17, \#FuerzaMexico,  \#Verificado19S, \#ayudaCDMX} that news reports related to the earthquake \cite{Tecnolog99:online,SISMOMXI46:online}, and @mentioned people who used such hashtags. 

\textbf{Facebook}. We posted an invitation to our survey on 48 groups that mentioned in their name or description one or more of the \#19S related hashtags or keywords that we have previously specified.

\textbf{Slack}. We posted invitations to our survey on the earthquake-related Slack channels of the group \emph{Codeando M\'{e}xico}. These channels all had the prefix \emph{``sismomx''} (e.g., ``earthquakemx'') followed by the channel's specific focus. For instance, ``\#sismomx-verificado19S'' was a channel focused on verifying \#19S information.

A total of 356 individuals responded to the survey. Of those 228 participants stated that they used technology during the relief efforts. Therefore, we conducted qualitative coding over their open-ended responses from those who reported using technology, to identify the different purposes that people had for using different social media in \#19S. We identified 8 categories (see Table \ref{table:helpCategories}). We hired three Spanish speaking college-educated crowd workers from Upwork to categorize open-ended survey responses. Two of them were requested to do the categorization, while we asked the third one to decide the final category for those responses in which the two coders could not agree on. The two coders agreed on 149 responses out of 185 (Cohen's kappa was 0.8; substantial agreement). We then asked the third coder to label the remaining survey responses upon which the first two coders disagreed. We used a ``majority rule'' approach to set the category of those remaining  responses.

Our survey revealed different \textit{purposes} that participants had for using multiple platforms. We dug deep into the most salient purpose which involved information verification. We conducted a content analysis on the digital traces left by participants to analyze their work dynamics around this purpose. Juxtaposing our survey's findings with the content analysis, helps us build up into a more macro-understanding of how participants worked to verify information using multiple platforms.

\subsection{Methods: Meso and Macro-Level} Our goal with the meso and macro-level analyses was to have a zoomed out at scale overview of  people's patterns for using different platforms. In specific, an overview of how information was assembled over time and the main groups per platform involved in content production and with the highest participation. We collected publicly available data from Facebook, Twitter, Slack, and GitHub related to \#19S from September 19 to October 19 of 2017. 

We chose to analyze Twitter, Facebook, Slack, and Github because these were the main platforms that the news media reported on and the ones that the survey participants most used. Github was mentioned by Slack users as the place where they were uploading the tools developed to aid the relief efforts; therefore, we decided that it was valuable to add in order to broaden our understanding of the collective efforts involved.

Table \ref{table:data} shows an overview of the data we collected for each platform.

\begin{table}[h]
\centering
\small
\begin{tabular}{ p{1.3cm} p{1.5cm}     p{2.3cm} p{2.2cm}}
{\bf Platform}& {\bf Participants} & {\bf  Posts} & {\bf Groupings}\\
\hline
Facebook & 10,262 & 12,606 posts & 48 groups and pages \\
Twitter & 792,665 & 2,909,109 tweets & 29,041 hashtags \\
Slack& 347 & 28,782 messages & 52 groups \\
GitHub & 216 & 2,602 commits & 26 repositories \\
\hline
\end{tabular}
\caption{\bf{ Data collected per platform}.}
\vspace{-1.5pc}
  \label{table:data}
\end{table}

Similar to \cite{wilson2018assembling}, for our meso-level analysis, we first plotted temporal graphs of information volume over time per platform. See Fig. \ref{fig:groupsperplatform}. This helps us to have a more general overview of how information was being generated day by day during the \#19S aftermath. By including temporal information in analyses of user activities, we can enhance understanding of how users navigate the information space, process information and make use of platforms \cite{SocialMe25:online}. Additionally, we analyzed which groups had the highest amount of participants and plotted how much content they were generating per platform. Our goal was to identify per social platform the groups that had the most number of members and also the most content production. For this purpose, we plotted for each platform and for each group in the platform the number of members the group had (X-axis) and the number of posts that the group generated (Y-axis), see Fig. \ref{fig:graphs}. For Twitter, previous work has used hashtags to detect groups of people with interests in common  \cite{silva2017methodology,bruns2012quantitative}. Therefore, we consider a group on Twitter as those tweets with a hashtag in common. For Facebook, we consider a group to be either a page or a group created. For Slack, we consider a group to be a Slack channel, and for Github, we consider a group to be a repository.

For our macro-level analysis, we conducted a network analysis to determine the participants who were considered ``influencers'' in each social network as well as those who could facilitate the ``spreading of information''. The measure that allows us to detect the ``influencers'' in a network is the betweenness centrality \cite{bokunewicz2017influencer}. The betweenness centrality measures the number of times a node acts as a bridge along the shortest path between two other nodes. A high betweenness centrality signals a strategic position within the network \cite{rusinowska2011social}.  To detect those participants that are able to "spread information", we use closeness centrality, as it estimates how easily a node
can reach other nodes in a network \cite{chen2012identifying,rusinowska2011social}. The nodes in the network are the accounts of people and organizations that were sharing information during the period studied. The links show how the information was flowing between the nodes. To detect clusters in the different social media networks we used the Louvain algorithm \cite{blondel2008fast}, a popular algorithm for community detection. In our case, we aimed at detecting clusters of users communicating among each other.

Our macro-level analysis helps us to understand broadly what grabbed people's attention on each platform (what content and what actors were important), and zoom out to obtain a broader picture of how the general public used at scale the different social media platforms. Here, we also study how people referenced content on other platforms. Then, we describe how we collected data from each platform:

\textbf{Twitter.} We used Gnip Power Track, a commercial data provider, to collect 2,909,109 tweets with the hashtags known to be used by people involved in the \#19S, these were the same hashtags we used to advertise our survey.
We operationalized ``posts'' as tweets, and ``groups'' as hashtags. The latter because previous research has shown that Twitter hashtags help cluster people around particular topics \cite{br2019corpus}, acting as ad hoc groups. 

\textbf{Facebook.} We used Facebook's public graph API to assess the creation of new groups and pages, and the posts people published on them. We tracked 48 groups and pages. We collected 12,606 posts from them. We asked permission to collect data from the administrators of the pages and groups. When possible, we also posted on the Facebook groups and pages to let people know we wanted to do an anonymized log analysis of their posts for research purposes.   

\textbf{Slack.} We collected 28,782 messages that people posted across the 52 channels related to \#19S that existed in \emph{Codeando M\'{e}xico's} Slack.  We counted the daily messages exchanged across all earthquake channels as ``posts,'' and the creation dates for the channels. We informed Slack administrators that we wanted to do an anonymized log analysis of their posts for research purposes.

\vspace{-0.1pc}
\textbf{GitHub.} Technologists used the platforms Slack and Github to operate\footnote{\url{https://github.com/CodeandoMexico/terremoto-cdmx}}. We first collected the GitHub links that people mentioned on Twitter, Facebook, and Slack in earthquake-related messages. Altogether, we identified 26 GitHub links pointing to different repositories, e.g., coding projects. We tracked their creation dates. Also, for each of these GitHub repositories we collected their description, IDs of people contributing to the project, and when commits occurred, e.g., changes to the source code.  In total, we collected 2,602 commits. We contacted the authors of each repository whenever possible to let them know that we wanted to do an anonymized analysis for research purposes. 

\section{Results}

\begin{table}
\small
\begin{tabular}{ p{3.4cm} p{1cm}p{6.2cm}}
{\bf \small Category} & {\bf \small Freq (Pct)} & {\bf \small Description}\\
\hline
\\
\small {1) Read, Share and Verify News} &  66 (29\%) & Any comment related to news reports about the earthquake. This includes reading and verifying news.\\\\

{\small 2) Share Needs}  & 52 (23\%) &  Share the needs of different parts of the country that were affected by \#19S.\\\\

3) Receive Updates from Friends and Family &  25 (11\%) & Use technology to confirm that family, friends, or co-workers are safe after the earthquake.\\\\

4) Donate Money & 25 (11\%) & Any comment related to donating either money, supplies or provisions to help \#19S victims.\\\\
5) Build Digital Tools  & 21 (9\%)& Create new technological tools to help the rescue efforts or to assist victims.\\\\

6) Conduct Data Analysis & 16 (7\%) & Perform data analysis or data processing.\\\\

7) Connect Strangers & 14 (6\%) & Connect people who did not know each other but who could collaborate and help each other in the earthquake aftermath. It also includes reconnecting people who went missing.\\\\

8) Mobilize People Offline  & 9 (4\%) &Incite people to physically go to locations to help earthquake victims or ask volunteers who are already on the ground to help in a particular way.\\\\

\hline
\end{tabular}
\caption{\bf{Description of the different purposes for which participants used multiple platforms in the \#19S aftermath. Information taken from 228 survey respondents who stated they helped the relief efforts using technology.}}
\label{table:helpCategories}
\end{table}
\subsection{Micro-level: Survey}

Our examination focused on how participants jointly used the platforms of Facebook, Twitter, Slack, and GitHub to contribute to earthquake relief efforts. The majority, 88\%(N=313), of survey responders were located in Mexico. The rest were participants in the Mexican diaspora. The median age of participants was 25 years old (SD = 9.8 years). When asked about how the earthquake affected them,  22\% (N=78) reported being personally affected, 39\%(N=139) had friends and family affected, 21\%(N=75) had some type of connection with earthquake victims, and 32\%(N=114) had no connection to anyone affected by the earthquake. Note that participants could have multiple connections to earthquake victims.  
Most participants in our survey, 64\% (N=228), stated they helped the relief efforts using  technology, investing a median of 20 hours: ranging from less than an hour to almost 3 months. Through our survey, we also identified that two additional platforms were used by participants during the earthquake: WhatsApp and Snapchat. While the information on Snapchat was not included in our investigation due to its potentially temporary nature and difficulty in validating the activity, we have investigated how specifically Whatsapp was utilized by participants in order to understand the information flow and use in coordination efforts during the aftermath of the \#19S earthquake \cite{simon2016kidnapping}.
Table \ref{table:helpCategories} presents a description of the different ways participants contributed to relief efforts using multiple technologies in response to our survey questions. 
Note that we only use the responses of participants who stated using technology (228 individuals). Therefore, our following analysis represents only the responses of those participants
\\


{\bf{Read, Share and Verify News.}}

The primary way in which participants felt they contributed to the relief efforts using multiple platforms, with 29\%(N=66) reporting this behavior, was by reading, sharing, and verifying news reports about \#19S. Participants considered it valuable to share news reports across social computing platforms because it enabled their different social circles to understand the important events occurring in the \#19S aftermath. Sharing particular news reports also seemed to help participants frame the events of the aftermath without creating a sense of panic:

\begin{quote} \emph{
``I tried to read through news reports and share useful information. If I did not do this, I knew that confusion and \textbf{panic would emerge} from all the information floating around. I tried to share news reports that could be useful...:P1''
} \end{quote}

Sharing news also usually included the activity of verifying the news before sharing it across multiple social computing platforms: 

\begin{quote} \emph{
``{I only shared the news that I had personally verified}. I didn't share chain news messages that I had no idea where they came from. So I shared very little [...] It was important to share verified news because when there's a lot of fake information circulating you only agitate...:P2'' 
} \end{quote}

Some individuals had strict norms about the type of news they shared. They adopted these norms to ensure that people knew what was really taking place across the country and could thus better identify where to help out:

\begin{quote} \textit{``We had a rule, if the news was true, then you needed to share photographs on Facebook and Twitter. This way we ensured that all the help we deployed was accurate...:P3''
} \end{quote}

Prior work had identified that people usually trusted citizen reporters who were physically in the disaster zone \cite{huang2015connected}. It is noteworthy that people in Mexico were skeptical of 
citizen reporters and actively had them verify their sources (e.g., by requesting photographs). We also saw participants start educating their peers about misinformation. Participants seemed to organize with friends on WhatsApp to curate together learning material. They then shared it on Twitter or Facebook to teach others about misinformation:

\begin{quote} \textit{``I verified news reports to foster a culture of education on the Internet. I wanted users to learn that they shouldn't fill common places, like Facebook, with misinformation [...] The key to our success was using WhatsApp and Google Docs. We discussed what educational message we wanted to communicate on WhatsApp. We then used the Google doc to actually create the content, all the details of what we would post on Facebook and Twitter.:P4''
} \end{quote}

Sharing news during natural disasters can lead to the emergence of misinformation \cite{zubiaga2018detection}. The correct handling of  misinformation is important as its spread can increase the sense of chaos in the aftermath of a crisis event \cite{alexander2014social}. Previous research had documented how in the past people have usually taken ``reactive strategies'' to deal with misinformation  \cite{starbird2014rumors,arif2017closer}. For instance, it was only after rumors spread that people reacted and tried to correct the postings. But for \#19S we saw people with more evolved and proactive attitudes, specifically verifying information to avoid creating rumors in the first place.

{\bf{Share Needs.}}
The second most common way in which participants contributed to the relief efforts was by sharing the needs that existed throughout Mexico in the aftermath of the earthquake, 23\%(N=52). Participants seemed to use multiple platforms because it facilitated accessing diverse social groups who could provide different perspectives about the current needs, such as location-specific needs or needs that only certain professions would understand, e.g., medical needs:

\begin{quote} \textit{``Facebook became a window of the needs of the city in specific points. It was very easy for any cyclist to be on a spot and upload something. It also helped us to reach different professionals, like musicians...:P5''
} \end{quote}

Needs tended to be shared with the following dynamic: 1) on-the-ground volunteers took notes about what was needed in different regions; 2) these individuals then shared this information via private messages, e.g., on WhatsApp, to online friends who then created Twitter or Facebook posts to inform the general population. 
Here, participants also struggled on the best ways to present needs without creating a sense of panic in others, but while inciting people to address those needs:

\begin{quote} \textit{``Virtually I communicated with my friends on WhatsApp to recompile and organize information about what donations were needed in different parts of the city. We organized a strategy on WhatsApp to inform Facebook volunteers about where they could bring their donations [...] We discussed what to post and how to post it to avoid creating panic. Others shared everything raw and that tended to incite fear.:P6''
} \end{quote}

{\bf{Receive Updates from Friends and Family.}}
One of the third most common ways (11\%(N=25)) participants used social media platforms to participate in \#19S relief efforts was by requesting friends and family members to provide status updates on their safety after the earthquake. The platforms participants selected to receive updates seemed to depend on the type of relationships they had:

\begin{quote} \textit{``I used my cellphone and landline to call and find the people I knew. A text message to the individuals I knew less. Email with people from my school who didn't answer the phone or did not have a Facebook profile.:P7'' 
} \end{quote}

There was a tendency to use multiple platforms to consecutively do the actions of receiving updates from friends and making \textit{donations}. Participants usually first communicated with contacts who were physically in the areas affected by the earthquake and then donated money to help relief efforts:

\begin{quote} \textit{``I got in contact with a friend on Whatsapp to know how he was [...] He was on the ground helping. I made a donation to him on PayPal and he bought food for earthquake victims...:P8''
} \end{quote}

{\bf{Build Technology.}}
Participants also used social media platforms during the earthquake to build technology (9\% (N=21)) and to conduct data analysis (7\% (N=16)). Note that all of these individuals considered themselves to be technologists, e.g., they had technological skills and worked in tech-related fields. Such individuals reported they primarily used Slack and GitHub to organize. These technologists reported that they deployed most of the tools they created on existing social computing platforms, such as Facebook or Twitter because they felt more individuals would likely use their tools if they co-existed in spaces people already used:

\begin{quote} \textit{``I participated in the Twitter bot project of \#FakeSismo that shared \#19S news [...] we moved to Twitter because users who were not experts like us use it frequently...:P9''
} \end{quote}

It was also interesting to observe that while technologists noted that their 
tools might not have been used by many \#19S victims, this did not seem to bother them. Their main goals seemed to have been more on investing in a culture of prevention:

\begin{quote} \textit{``More than just creating tools to analyze what happened, it's about creating prevention [...] generating conscience and a culture of prevention...:P10'' 
} \end{quote}

The culture of prevention in the event of a crisis could be considered dual-purpose: building technology to coordinate rescue efforts and implementing tools to assist with the flow of information. Instituting a process and a mentality of assisting rescue efforts in situations of crisis might be the obvious result in creating a culture of prevention; however, we can also understand their efforts to promote tools that permitted the flow of information as a necessary prevention against misinformation solidifying in people's minds during crises \cite{lewandowsky2013misinformation}. Their efforts thus set a precedent for facilitating the ability to refute misinformation during future crises.

{\bf{Connect Strangers.}}
Similar to previous findings \cite{reuter2018fifteen}, participants also used technology for connecting two or more strangers so that they could help each other in the aftermath of \#19S (6\%(N=14)).  
Participants tended to read in Facebook or Twitter the needs that existed in Mexico, and then actively searched across social computing platforms to find people who could address those needs:
\begin{quote} \textit{``I read people's needs on Twitter and then I would connect them with people who I thought could help them (doctors I knew, vets, psychologists, translators.):P11''} 
\end{quote}

{\bf{Mobilize People Offline.}} 
Participants also used multiple platforms to incite people to physically go to certain locations and help earthquake relief efforts (4\%(N=9). Mobilize people offline included organizing volunteers who were already currently on the ground to help in a particular way. We observed that the activities of ``Sharing Needs'' ``Reading, Sharing and Verifying News'' and ``Mobilize people offline'' appeared to be closely related. Individuals tended to use Twitter to sort through the needs and news reports to understand what was taking place and where help might be the most needed. They would then go on Facebook, Messenger or WhatsApp to convince and mobilize their contacts to take offline action and help:

\begin{quote} \textit{``Twitter helped me to identify what was going on in each collection center. I would then go on Facebook and message my students. My hope was that I could motivate them to go to a particular collection center to help...:P12''
} \end{quote}

The individuals who engaged in this activity appeared to be those who considered they did not have the physical condition to help in offline relief efforts. But they found this online activity as a way to positively contribute. This resembles how people with disabilities contribute to offline activism \cite{li2018slacktivists}. 

\begin{quote} \textit{``My wife was the one who advised me, that although I may be old I could help coordinate those volunteers on the ground by telling them how to follow security measures, how to evacuate a building safely, where to get support, and also where their help was most needed...:P13''
} \end{quote}

It is worth noting that the percentage of this category is below  ``Conduct data analysis'' or ``Build digital tools''. This could be due to characteristics of individuals who answered our survey. It is highly likely that participants who spend more time on Slack were more willing to answer our survey than those who were working offline on relief efforts. 

\subsection{Micro-level: Content Analysis}
Our survey revealed that one of the main reasons why participants used multiple social media platforms in the \#19S aftermath was to read, share and verify news. We, therefore, aimed to understand this process more deeply. For this purpose, we adopted an approach similar to \cite{wilson2018assembling} and examined the digital traces (conversations, hashtags, tweets, linked content and websites referenced) left by participants on the different platforms they reported. Analyzing these observations, helps us build up into a more macro understanding of how people worked to support an information verification process during the earthquake.

\begin{figure}
  \begin{center}
    \includegraphics[width=0.9\linewidth]{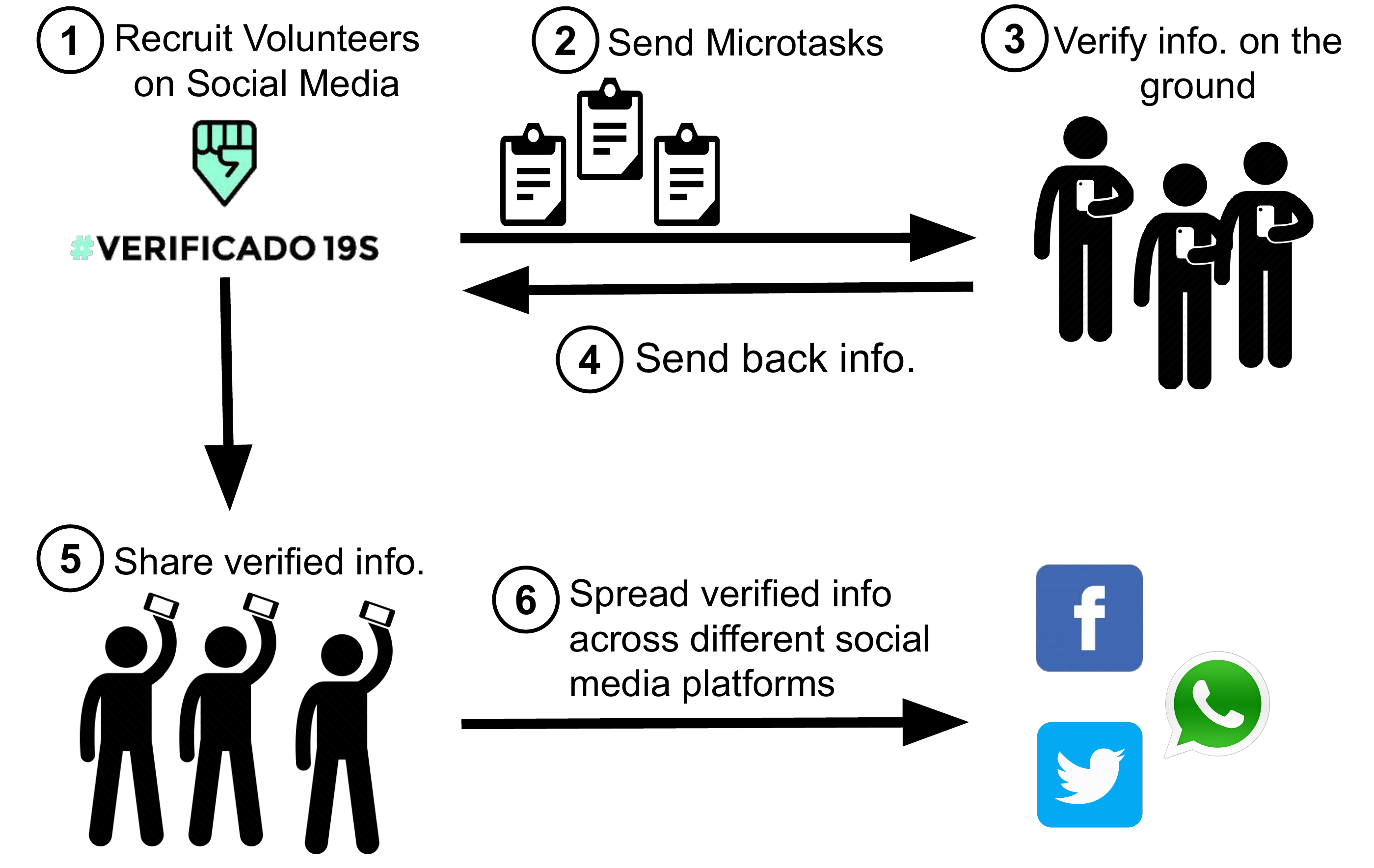}
  \end{center}
\vspace{-1pc}
  \caption{\bf Crowdsourcing Information Validation}
\label{fig:InformationValidation}
\end{figure}

\textit{\bf{Crowdsourcing Information Validation}}.
We realized from people's Twitter posts, that in order to collect and verify information, they were using a crowdsourcing mechanism where they gave people micro-tasks to help in the verification process, see Fig. \ref{fig:InformationValidation}. 

\begin{wrapfigure}{h}{0.4\textwidth} 
 \vspace{-3.5pc}
  \begin{center}
    \includegraphics[width=0.4\textwidth]{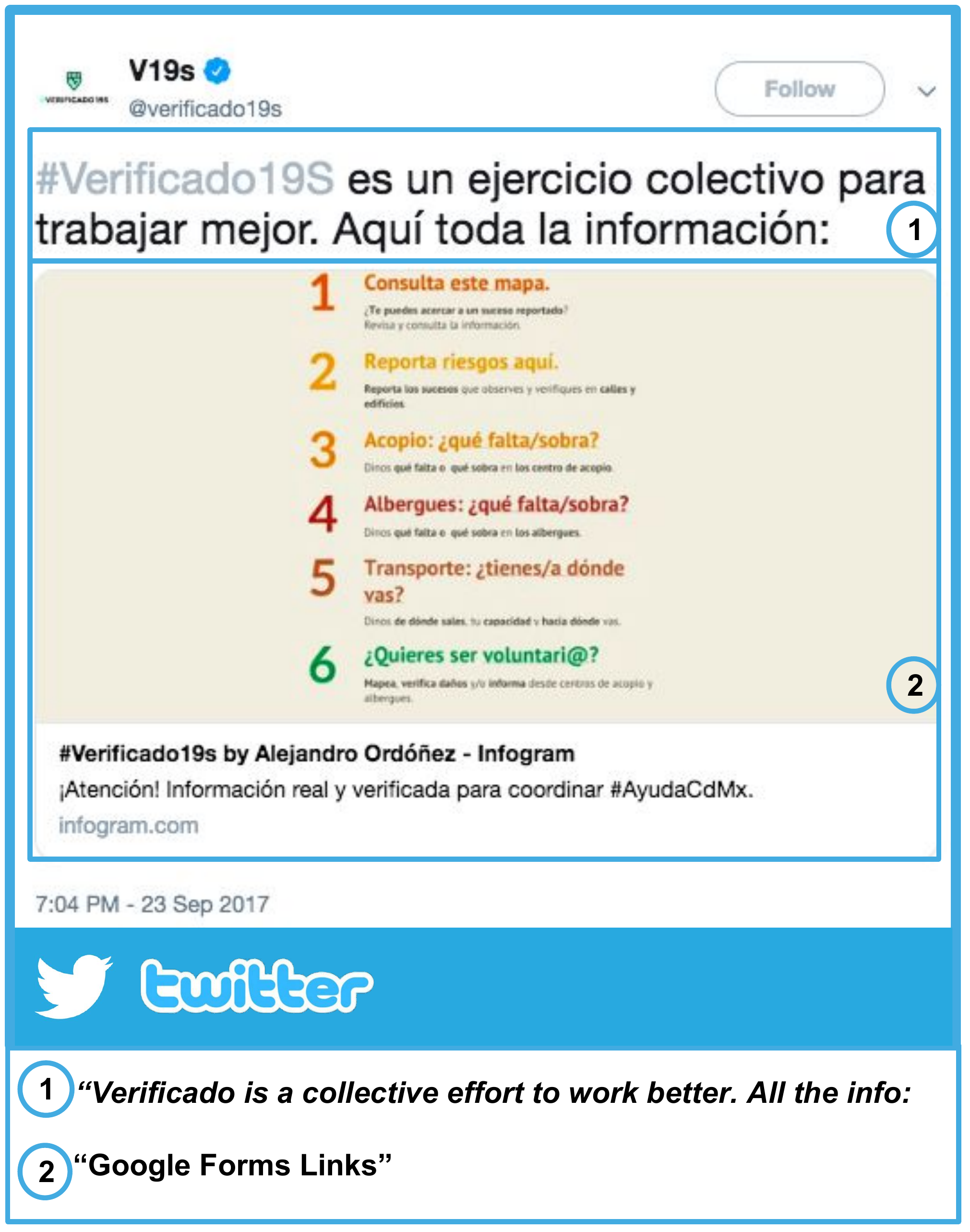}
  \end{center}
  \caption{\small{\bf{Tweet with a link to a Google Form requesting people to state how they wanted to help out in the relief efforts.}}}
  \label{fig:TwiiterVerificadoCrowd}
\end{wrapfigure}
They first shared a tweet that included a link to a Google Form. They requested people to fill out the form and state how they wanted to help out in the relief efforts. Some of the options people could select included: digital volunteering opportunities such as input verified information to online maps, verify damages and needs for collection centers and shelters; report damaged buildings/streets and verify citizen news reports; and transport and delivery of materials or people in need, among other options. Fig. \ref{fig:TwiiterVerificadoCrowd} presents an example of such type of tweets. Depending on what people had selected, they were then tweeted specific micro-tasks to start helping (e.g., ``help verify that a shelter needed water'').

These mechanisms highlight how people utilized user-generated content not only to find out necessities on affected areas but to fact-check word of mouth reports about the situation in the city. People in these cases seemed to have worked closely with \emph{@Verificado19S} as they mentioned the account frequently in their tweets and retweeted their content.  \emph{Verificado19S} appears to have functioned as a civic response group that brought trust and through this trust could crowdsource and organize on-demand volunteers on the ground to verify information \cite{AfterMex8:online}. Previous studies have found that if information providers clearly identify themselves and share their goals of giving reliable information, it grants them credibility and facilitates the co-participation of citizens  \cite{chauhan2017providing}. This dynamic appears to be what we are identifying with \emph{Verificado19S} in the \#19S. 

\textit{\bf{Dealing with Outdated Information}}. Through our manual Twitter analysis, we identified that \emph{Verificado19S} started standardizing all its \textit{calls to action} (e.g., posts that asked citizens to volunteer in a specific way.)

\begin{figure}
  \begin{center}
    \includegraphics[width=0.65\linewidth]{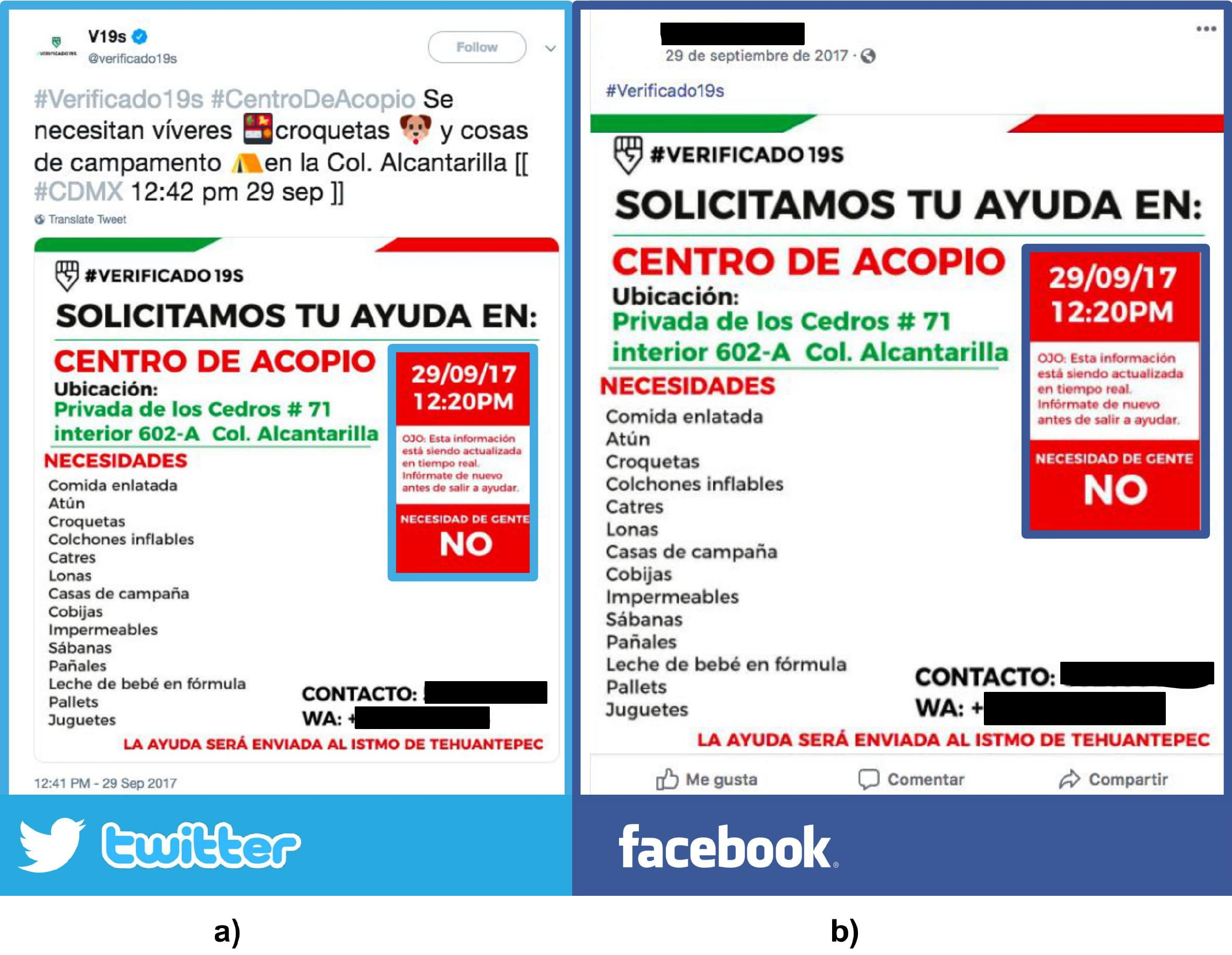}
  \end{center}
\vspace{-1pc}
  \caption{\bf{a) Tweet with a call to action from the \emph{@Verificado19S}'s Twitter account. b) Facebook post with an image taken from \textit{@Verificado19S}'s Twitter feed.}}
\label{fig:TweetVerificado19s}
\end{figure}

\emph{Verificado19S} used a template that involved an image with the date, time, place, specific needs and contact person, as well as the place where things were needed, see Fig. 
\ref{fig:TweetVerificado19s} a). \emph{Verificado19S} constantly tweeted multiple times per day about resources that were needed in different disaster zones and shelters. We believe that by integrating a date and time to the images, it directly helped citizens to identify what information in a given time frame was relevant. Therefore, if the same image was shared within other social platforms later (see e.g.,  Fig.\ref{fig:TweetVerificado19s} b), people could  better decide on whether they wanted to take action or not (e.g., it might have been something that was needed days ago, so it might not be urgent anymore.) 

\begin{figure}
 \begin{minipage}{0.47\textwidth}
  \begin{center}
    \includegraphics[width=1\textwidth]{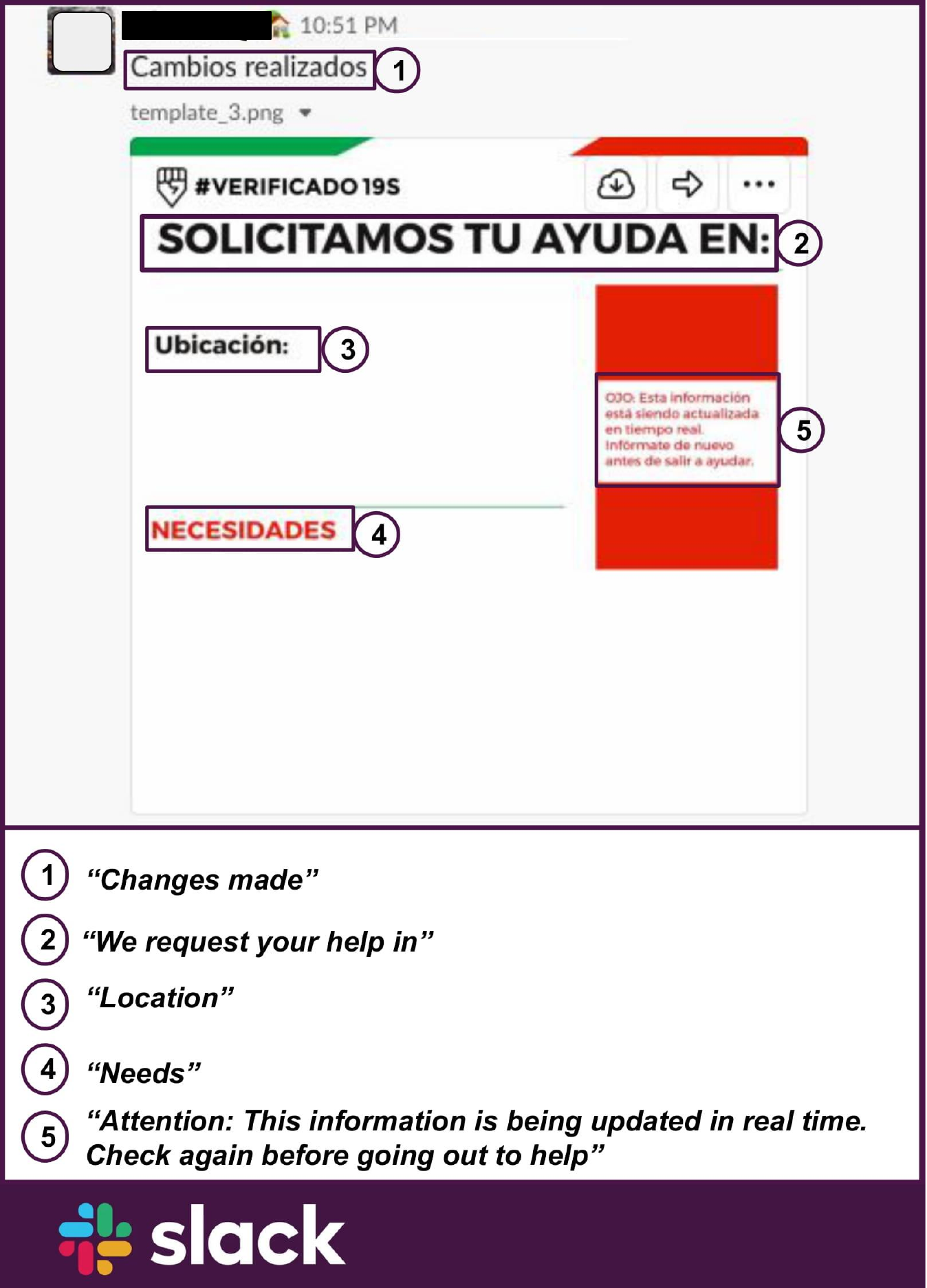}
  \end{center}
  \caption{\small{\bf{Template created by the Slack crowd to be used by \textit{@Verificado19S}}}}
  \label{fig:SlackTemplate}
  \vspace{-1pc}
\end{minipage}\hfill
\begin{minipage}{0.47\textwidth}
  \begin{center}
    \includegraphics[width=1\textwidth]{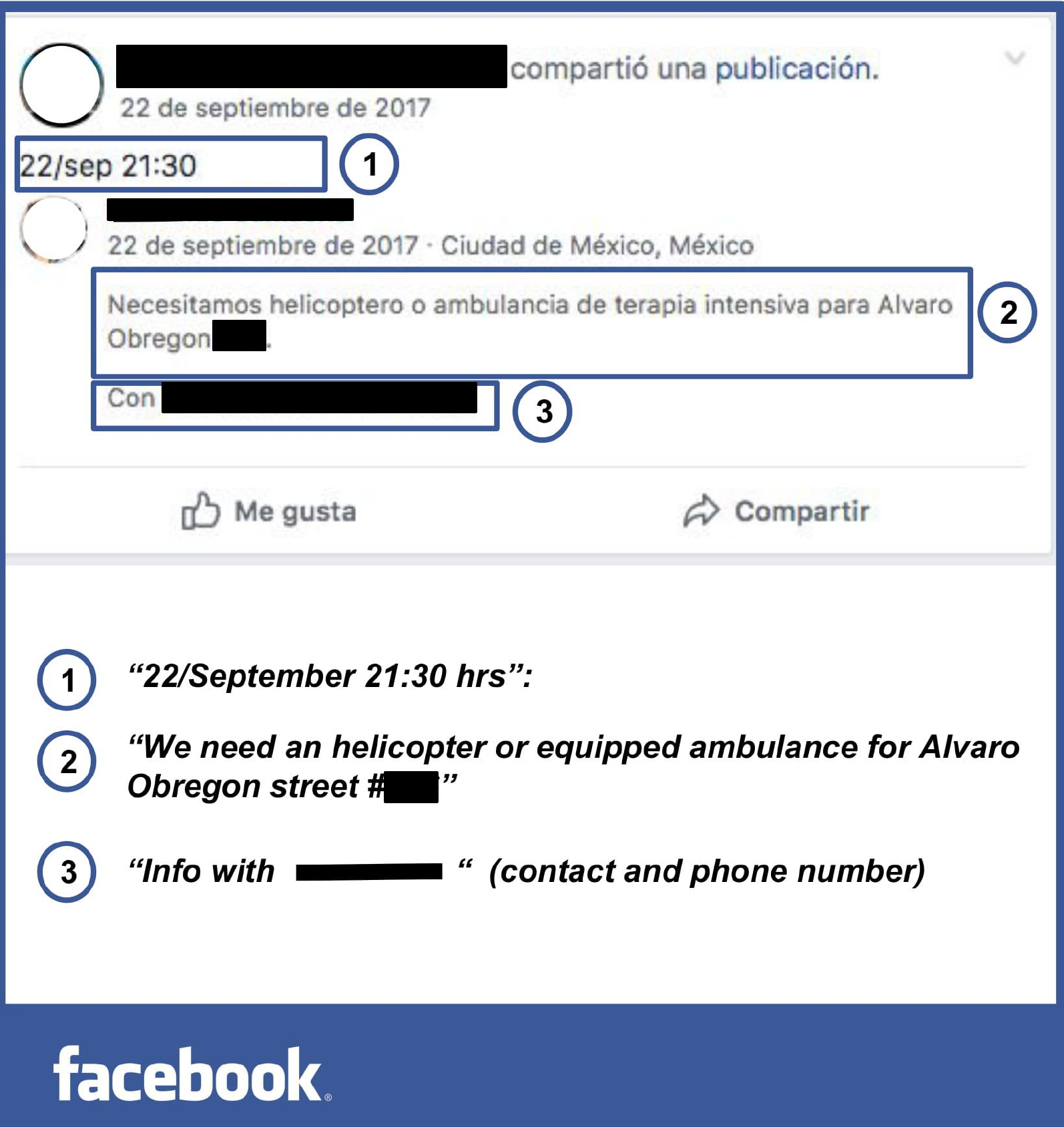}
  \end{center}
  \caption{\small{\bf{Facebook post reshared by a user adding date and time.}}
  \vspace{-1.5pc}
  \label{fig:FBUserstimedate}}
\end{minipage}
\end{figure}

 
Our Slack analysis revealed that citizens operating within Slack had been the ones to brainstorm ideas that ultimately led to developing the image template that was later adopted by \emph{Verificado19S} in order to share needs at scale. Fig \ref{fig:SlackTemplate}. shows an example of the template that participants were collectively designing on Slack. We also analyzed the Facebook posts that our survey participants shared within public Facebook groups. We found that people started adopting some of the same conventions that \emph{Verificado19S} had been pushing. Fig. \ref{fig:FBUserstimedate}, shows how a participant of a Facebook group reshared a post, adding the approximate date and time it was posted. This is noteworthy as past research has suggested that information that is outdated can affect the coordination and efficiency of relief efforts \cite{wong2017social}. It is interesting to observe how people in the \#19S earthquake were able to develop their own mechanisms to overcome this problem across social media platforms. 

\textit{\bf{Educating People about Misinformation}}. By manually analyzing the online interactions of our survey participants on Slack, we found that participants were organizing to develop tools to deal with misinformation during the crisis. Some of the tools included: 1) BanFakeNews, a web platform to report fake news being spread during the crisis, 2) Fake News Bot, a Twitter bot to distribute and classify information on social media about possible fake news. Our digital trace analysis led us to find the first messages in which participants discussed organizing to tackle the problem:

\begin{quote}
 \emph{This is a big problem [misinformation] because on social media there are a lot of people sharing tons of it.: S\_user1}   
\end{quote}

\begin{figure}
  \begin{center}
    \hspace{0pc}{\includegraphics[width=0.7\linewidth]{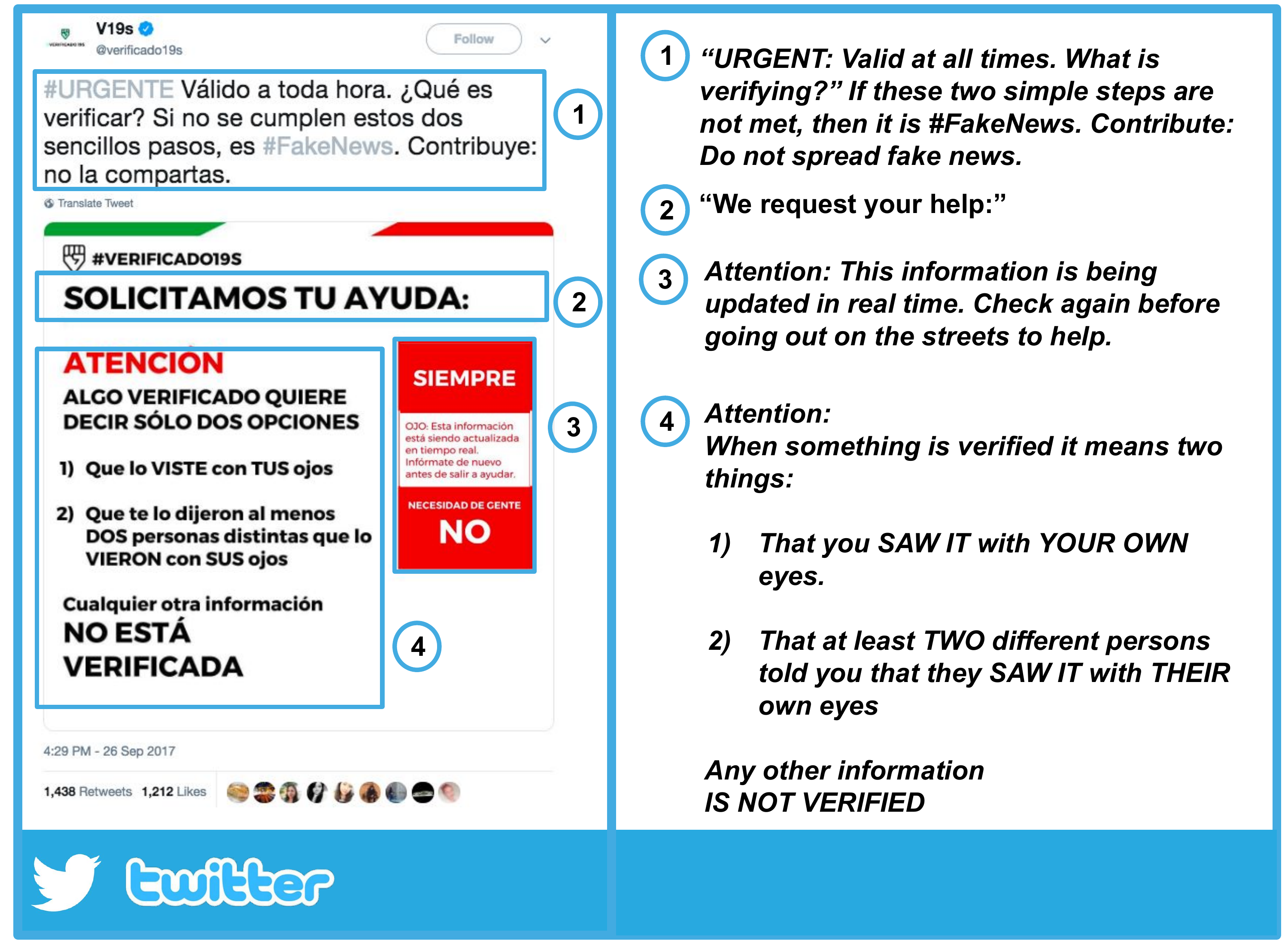}}
  \end{center}
\vspace{-0.5pc}
  \caption{\bf{Image created to educate users about misinformation: Attention: Something Verified means only two things: 1) That you saw it with your own eyes. 2) That at least two different people saw it with their own eyes. Anything else is not verified.}
\label{fig:MisinfoVerificado}}
\vspace{-0.8pc}
\end{figure}

Someone suggested creating an exclusive Slack channel to focus on dealing with the problem:
\begin{quote}
 \emph{Create a channel to focus on the problem, as it is a big one. Make a request to (S\_Volunteer1) to create it. It could be called. sismomx-fakenews [earthquakemx-fakenews].: S\_user3}   
\end{quote}
On the newly created channel  called ``sismomx-fakenews'', participants  started to brainstorm ideas about how to solve the problem of misinformation. After some discussion, they decided to develop an educational flyer that could educate people about misinformation and how to easily verify anything they wanted to share online. This flyer was also shared by \emph{Verificado19S} to educate people at scale about misinformation, see Fig.\ref{fig:MisinfoVerificado}. This image was distributed by the Twitter account of \emph{Verificado19S} and obtained 1,711 retweets and 1,266 likes.

\subsection{Results: Meso-Level} 
Our meso-level analysis studies how information was assembled on different social media platforms. This level allows us to view how information was being generated. For this purpose, for each platform, we study: 1) the number of daily posts; 2) groups that had the most number of members and the most content production; 3) references to other platforms.

\subsubsection{Daily number of posts.} 
\begin{figure}[t]
\centering
\includegraphics[width=1\textwidth]{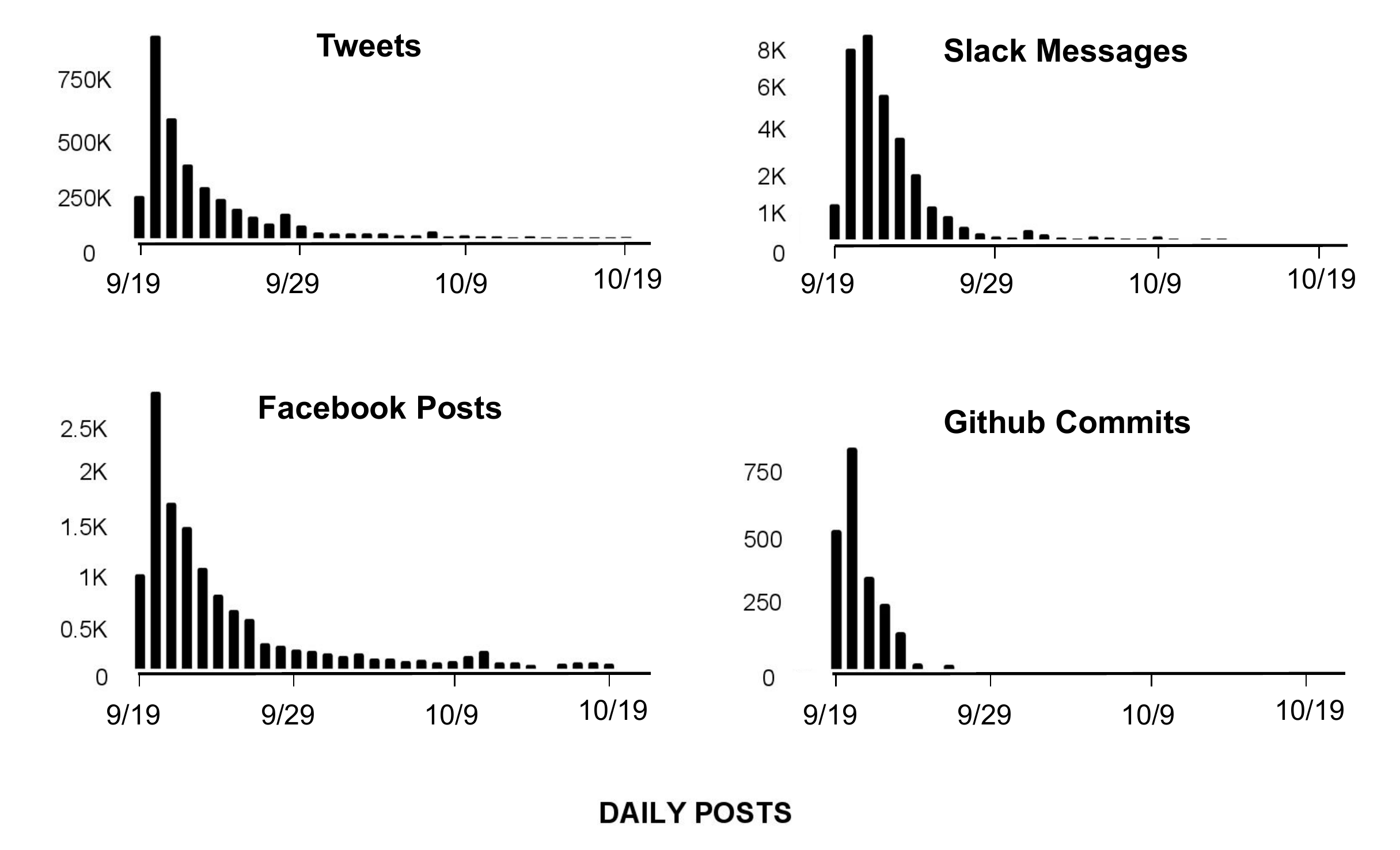}
 \vspace{-0.5em}
\caption{\bf{``Posts'' across platforms through time.}}
\label{fig:groupsperplatform}
\end{figure}

Fig. \ref{fig:groupsperplatform} presents the daily number of posts generated per platform over time. We notice that for all platforms, the number of posts decreased overtime. Twitter was the platform where people produced the largest content volume that we could observe from our data. The highest peak occurred on September 20 of 2017, the day right after the earthquake,  with  865,508 tweets. There is an interesting outbreak on October 7 that occurred because the national football team paid tribute to Frida, the dog who became famous for her assistance in the earthquake relief efforts \cite{Fridathe41:online} and a symbol of hope in the country \cite{Fridadog1:online}. The tribute was covered by international media \cite{Fridadog1:online} and people were tweeting about it.  

The second platform where people produced the most content was Slack. However, the highest peak on Slack occurred on September 21 of 2017. We hypothesize that this might have occurred because technical people took the time to organize and recruit others to build tools for the earthquake response efforts.  Additionally, we observe that across all platforms the activity quickly died out as the days advanced. It is interesting to note, however, that on Slack on October 1st, there was a brief reactivation; people suddenly started posting again more. Upon inspection, we observed that on that date there had been an aftershock from \#19S. This event might have driven people's reactivation. The reactivation only happened on Slack (where primarily the technologists operated.). 
On Facebook, there was a reactivation on October 11th. After further inspection of the posts, we did not find evidence of anything unusual shared on the platform during that day. The posts were primarily from people within the Facebook groups that aimed to reunite lost pets with their owners.

In general, all the graphs decreased in activity as the days passed. Research that analyzed media coverage on Twitter from the same  event \cite{curiel2019temporal}, found that the tweets related to the earthquake had an initial peak and then an exponential decay as the days passed. We found that this was not only occurring on Twitter but it also happened on the other social networks as well.

\begin{figure}[t]
\centering
\includegraphics[width=1\textwidth]{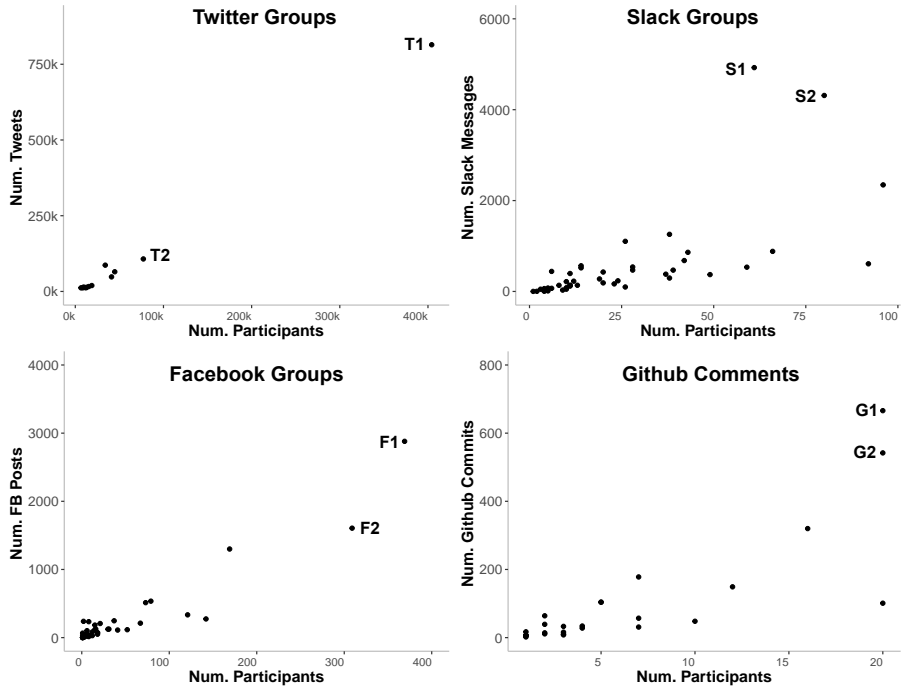}
    \caption{\bf{Content per ``group'' on each platform.}}
\label{fig:graphs}
\end{figure}

\textbf{\bf{Groups with the highest amount of participants and content.}} In our meso-level analysis, we identified which group had the highest amount of participants and content, see Fig.\ref{fig:graphs}. This helps us to understand which ``groups'' got the most amount of attention.

{\bf Twitter.} The hashtag with the most number of tweets was  \#FuerzaMexico (point T1), (it  means \textit{Mexico Strong}), see Fig. \ref{fig:graphs}. This hashtag helped bring attention to the aftermath of the earthquake. The hashtag was used to share ways to help \cite{FuerzaMe54:online}. The hashtag had 814,340 tweets with 404,218 participants. The second hashtag with the most amount of tweets was \#Verificado19S. It had 107,280 tweets with 77,209 participants.  

Previous work \cite{comunello2018prelims} found that  a  hashtag has the ability to draw large audiences, but its widespread usage during a crisis event  can create inefficiencies in communicating relevant information. As the information is generated  by thousands  of participants at a high speed, this could create a significant hurdle to find relevant content. 
We found that in some cases this was happening, as requests for help were tweeted using more than one hashtag.

{\bf Facebook.} The group with the most number of participants and posts had the purpose of  connecting lost pets with their owners (point F1 in  Fig. \ref{fig:graphs}). The group had 369 participants that generated 2,880 posts in total.  The second group with the most number of posts (point F2) had 1,606 posts in total and 309 participants. It was related to  ``sharing needs'' and ``connecting strangers''. The fact that the group with the most posts was related to reuniting pets with their owners, and the second one belonged to categories in which people were requesting help and connecting strangers, tells us about the importance of having reliable information. We believe that mechanisms such as the one developed by \emph{Verificado19S} were important to  make relief efforts more efficient. 

{\bf Slack.} For Slack, We discovered that the most active channels were those related to ``building digital tools'' and ``building websites.''. The channel with the most messages was ``sismomx-fakenews'' (point S1 in Fig. \ref{fig:graphs}). As mentioned previously, this channel focused on creating a website to report fake news. It had 4,926 messages and 61 participants. The second channel with the most messages was ``sismomx-acopio-api'', with 4,314 messages and 80 participants (point S2). This channel focused on creating APIs to query data about available shelters. This empowered others to do data analysis or build tools (e.g., people built a website\footnote{\tiny\url{https://www.sismosmexico.org/}} that used the API to provide real-time information about the shelters).

\textbf{Github.}  Similar to Slack we decided to analyze the repositories with the greater number of participants and commits (changes to the project) on Github. With this, we wanted to uncover what repositories where considered more relevant. Studying these variables helps us to identify the projects where people were focusing their attention on as they were contributing code and making updates. Fig. \ref{fig:graphs} point G1, shows that the repository with most commits (666)  was for building a website where people could easily access all the information about shelters, collection centers, hospitals, etc. It had 20 participants in total. The second repository with most commits (542) was a tool to  provide real-time information about the shelters (point G2). 

\subsubsection{Cross-references between platforms.}
We analyzed how people cross-referenced platforms to get a perspective of how the users of a particular platform might amplify information from other online spaces. Table \ref{table:mentions} shows a comparison of the mentions of different platforms within a specific  online space. We see that people on Facebook did reference Twitter, but there was no mention of Slack or GitHub. Similarly, people on Twitter did reference Facebook but made little or no mentions of GitHub or Slack. People on Slack, on the other hand, did reference all the other platforms with Twitter and GitHub being more prominent. This makes sense when we recall that in our micro-level analysis technologists organized their tool building within Slack and Github, and they built tools that coexisted on Facebook or Twitter. Slack appeared to have acted in some cases as a backstage into what was shared and posted on other social media sites, such as Twitter. 

\begin{table}[h!]
\small
\centering
\begin{tabular}{|l c c c c|}
\hline
{\bf } & {\bf Facebook} & {\bf Twitter} & {\bf Slack} & {\bf GitHub}\\
\hline
{\bf Facebook} & - & 19 & 0 & 0 \\
{\bf Twitter} & 2115 & - & 0 & 1\\
{\bf Slack} & 70 & 287 & - & 766\\
{\bf GitHub} & 7 & 25 & 17 & -\\
\hline
\end{tabular}
\caption{\bf{Number of cross-references that people within a particular platform made about other social computing platforms.}}
  \label{table:mentions}
  \vspace{-1.5pc}
\end{table}

\subsection{Results: Macro-Level}\label{RESULTS: MACRO LEVEL}

To obtain a broad overview of who was involved in the information space within each social media platform, we analyzed the  structure of the conversations on each platform through network graphs. For this purpose, we conducted a network analysis similar to what prior research in crisis informatics has used to conduct a macro-level type analysis \cite{wilson2018assembling}. To further understand how these different actors interacted with each other, we adopted an approach similar to \cite{arif2018acting} and inspect the interactions of the accounts using interpretative analysis of a network graph based on graph flows.

For the network analysis, we created network connections between two users depending on various ways in which they interacted: on Twitter, it was based on whether they reshared each others' tweets; on Facebook, it was based on whether a user posted a message on other users post;  on Slack, it was based on whether they replied (mentioned) each other in a message; on GitHub, it was based on whether they contributed to the same projects together or forked each others' projects, e.g., they made commits to the same GitHub repositories or made copies of each other's project. We used the Louvain algorithm to detect clusters of users communicating among each
other, and the ForceAtlas2 algorithm to obtain the overall structure of the network graph \cite{jacomy2014forceatlas2}.

\begin{figure}[t]
\centering
\includegraphics[width=1\textwidth]{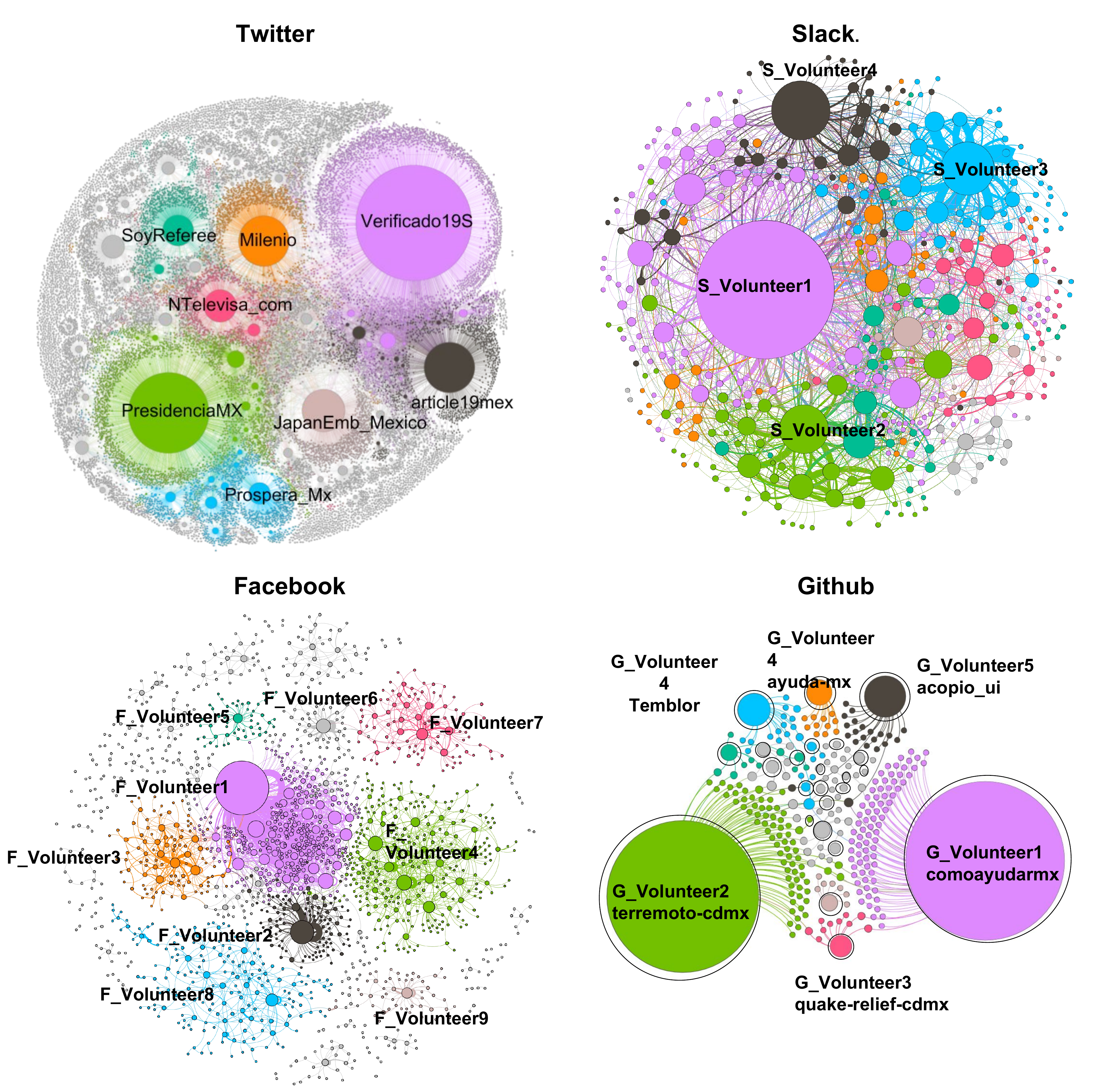}
    \caption{\bf{Network Graphs per Social Computing Platform.}}
\label{fig:earthquakegraphs}
\end{figure}

In Fig. \ref{fig:earthquakegraphs}, each node represents a particular account, sized by the number of interactions with the other accounts. For example, for the Twitter graph, large nodes represent a highly reshared accounts. Nodes are connected via a directional edge from a resharing account to a reshared account. Each edge is weighted by the number of reshares between the two. 

For each graph, we used the Louvain algorithm to uncover and color-code the different networks (clusters) present in each social computing platform. For each platform, we analyzed the main clusters of each network and described the most relevant actors within the information space. For each graph, we calculate the betweenness centrality and the closeness centrality of the nodes. The betweenness centrality allows us to detect the ``influencers'' of a network as nodes with high betweenness facilitate the information flow. The closeness centrality allows us to detect the ``spread of information'' through the network, as it allows us to measure how easily a node can reach other nodes in the network.

{\bf{Twitter Network Graph.}} Fig. \ref{fig:earthquakegraphs} presents the network graph we computed for Twitter. Here, we iteratively visualized the network graph (and hence the information flow) by defining nodes to be Twitter accounts and directed edges to be retweets between the accounts.  

We see that the ``influencers'' in the network are \textit{@PresidenciaMX} and  \textit{@Verificado19S} with betweenness centrality  value of 1,468,135 and 1,051,938  respectively. By looking at their closeness centrality, we can see that they are both similar, \textit{@PresidenciaMX} has a value of 0.3367 and \textit{@Verificado19S} has a value of 0.3124. However, even though both are influencers in the network with a high capacity of spreading information, there was no communication between the accounts.  The most retweeted account was not an account from an official long-standing institution in Mexico, but 
from \textit{@Verificado19S} \cite{HowtheVe71:online}. This account was created on September 23rd, 2017 (four days after \#19S). \textit{@Verificado19S} was retweeted a total of 17,638 times during the period collected. Overall, \textit{@Verificado19S} seemed to have operated within the category of ``Reading, Sharing, and Verifying News'' that we discovered in our micro-level analysis. 

The second most retweeted account belonged to Mexico's presidency \\\textit(@PresidenciaMX). Its tweets were retweeted a total of 15,168 times (2,470 times less than \textit{@Verificado19S}). Upon manual inspection, we identified that the tweets from Mexico's presidency seemed to have focused on providing tips to recognize misinformation, and request that people complete a survey of their current status (i.e., were people OK?). Overall, \textit{@PresidenciaMX} seemed to have operated within the ``Reading, Sharing, and Verifying News'' and ``Receiving Updates'' categories. 

From the node structure of Fig. \ref{fig:earthquakegraphs}, we observed that the accounts of \textit{@Verificado19S} and \textit{@article19mex}, an NGO fighting for people's free speech and information rights, had overlap in the nodes around them. This means that these accounts tended to retweet and amplify each others' content. This result makes sense as @article19mex belongs to the conglomerate of organizations that cooperated with the \textit{@Verificado19S} effort \cite{Verifica83:online}. We notice similar patterns in the Mexican government accounts: \textit{@PresidenciaMX, @Prospera\_Mx}. It is interesting that the accounts focused on information verification (\textit{@Verificado19S} and \textit{@article19mex}) seemed to have operated independently from the Mexican government;  and that the news media (\textit{@Milenio, NTelevisa\_com, @SoyReferee}) appeared to have acted as an intermediary between these two types of organizations (amplifying both their messages). We believe that this was due to the lack of trust in the government that has been an issue in the overall region \cite{bargsted2017political}.  

{\bf{Facebook Network Graph.}} Fig. \ref{fig:earthquakegraphs} presents the network graph we computed for Facebook. We iteratively visualized conversation flows between accounts by constructing the network graph in which we defined nodes to be Facebook accounts and directed edges to be messages. In this graph, we see that one of the main differences between the network graphs of Facebook and Twitter is that on Facebook the accounts with the most influence were not accounts from celebrities, but rather Facebook Groups created by ordinary people; this is, people who were Facebook group administrators. The account with the highest betweenness centrality had a value of 181,488 and a closeness centrality of 0.2908. These accounts also appeared to operate more within the category of  ``Connecting Strangers'' that we identified in our micro-level analysis.

The accounts whose posts were commented the most came primarily from Facebook groups  related to lost and found pets. Additionally, these groups tried to connect people who had found a lost pet with the pet's owner. The slight separation that exists between clusters in Fig. \ref{fig:earthquakegraphs} is because people appeared to have operated more within particular Facebook groups that managed very local information, e.g., F\_Volunteer7 shared only photos of lost dogs from a specific neighborhood in Mexico City (note that we anonymized names of ordinary people's accounts).

{\bf{Slack Network Graph.}} Slack does not have a reshare button. To draw the network graph we utilized information about how much each account was mentioned by others. We visualized mention flows between accounts by defining nodes to be Slack accounts and directed edges between the nodes to be mentions between accounts. The Slack network graph in Fig. \ref{fig:earthquakegraphs} reveals a large central cluster, in purple, of the account with the highest betweenness centrality: \textit{S\_Volunteer1}. Upon manual inspection, we identified that \textit{S\_Volunteer1} was the director of \emph{Codeando M\'{e}xico} who tried to organize everyone to build digital tools and was organizing most of the work on the different Slack channels (hence why he was mentioned the most). The betweenness centrality of the account (influence power) was 37,255 and his closeness centrality was 0.6306. The account was mentioned a total of 937 times by 177 users. The other most mentioned accounts were accounts leading the production of different tools (in the description of the Slack channel where they operated they were usually also mentioned as the leaders). 

{\bf{Github Network Graph.}} GitHub does not have a direct reshare button. Thus, we analyzed those users whose GitHub repositories had the largest number of contributors and the most number of individuals who forked their project. We created a network between two people if one person had directly participated in the GitHub repository of the other or they had forked their project. From Fig. \ref{fig:earthquakegraphs} we see that only two volunteers with their repositories were the ones with higher betweenness centrality: the volunteer who created the repository of \emph{``comoayudarmx''}, with a betweenness centrality of 3,376 and a closeness centrality of 0.3608., and the volunteer who created the repository of \emph{``terremoto-cdmx''}, with a betweenness centrality of 3,463 and a closeness centrality of 0.3559. The repository of \emph{comoayudarmx} had 86 contributors with an average of 8 commits per contributor and a total of 666 commits.    Similarly, \emph{terremoto-cdmx} had 82 contributors, with an average of 7 commits per contributor and a total of 542 commits. Other repositories had only one contributor. The main characteristic that appeared to exist between these two Github volunteers is that their repositories were also the most mentioned on Slack.




\section{Discussion}

In this section, we explain  different ways in which people adapted their use of social media to overcome its limitations. We compare our findings  with the previous theoretical understanding of behavior during crisis events and use them to make recommendations for technology design in order to improve the effectiveness of social media use during crisis response efforts.

\subsection{Misinformation Mitigation During Crisis Events}

Past research \cite{gupta20131,gupta2013faking}, has identified that during crisis events due to heightened anxiety and emotional vulnerabilities, people are often more susceptible to believe and share misinformation. In another study, Starbird et. al \cite{starbird2014rumors} found that during disasters individuals usually are more ``reactive'' with misinformation. They wait until it is a problem to react, correct, and stop it. However, corrections tended to spread more slowly than misinformation. Prior work has also shown that social computing platforms facilitate constructing an infrastructure for crisis response \cite{dailey2017social}. This is, users are not confined to use just one social media platform during a crisis event; they use a plethora of them to be informed about the unfolding events that occur in the aftermath of a disaster. However, they still find barriers to using social media during a crisis situation because of a lack of trust in the information they see \cite{lee2019explicit}.  Due to the fast paced nature of information circulating during a disaster event, preventing the spread of misinformation in the first place is important.

Our research uncovered different mechanisms that can help mitigate the spread of misinformation during a crisis event. The first one was an organized citizen-driven approach. We found that \textit{Verificado19S} acted as an independent and trustworthy organization that used on-the-ground citizens to verify information. In this way, \textit{Verificado19S} was able to engage and organize citizens in the process of information verification, e.g., organizing people who were within certain streets to help verify specific disaster information.  

\subsection{Keeping Information Relevant} 

The second approach was dealing with outdated information. Previous work stressed social media's ability to help citizens self-regulate inaccurate and outdated information \cite{soden2016infrastructure}, but it has been found that self-regulation does not happen at the pace needed for supporting logistic efforts during a crisis \cite{wong2017social}. 

Past work \cite{hiltz2013dealing,starbird2012will} has identified that  ensuring that the information that is shared online is ``relevant'' and ``trustworthy'' are some of the major difficulties that people encounter when using social media during disasters. Information that is outdated (and hence no longer relevant) can  complicate response efforts, jeopardizing the safety of first responders and the community \cite{wong2017social}. 

To overcome this, \emph{Verificado19S} developed a mechanism to keep important information  useful and understandable over time, even if it was spread on different social media platforms.  Our analysis uncovered how \emph{Verificado19S} developed its own organic mechanisms to start to bring a sense of ephemerality to their posts and overcome the fact that most social media platforms operate with data permanence. In specific, \emph{Verificado19S} used simple mechanisms, such as encoding timestamps into images to give people a sense of what information was outdated and which one was relevant. This lead to a form of \textit{forgetful digital memory}, which ensured that the important information was kept useful and understandable over time \cite{niederee2015forgetful}. 

This capability can be replicated by social media platforms during times of crises to give more visibility to time-sensitive information as the crisis is unfolding, and keep information that is shared on different social media platforms relevant. For instance, if someone wanted to reshare a need that was requested a long time ago, the platform could create mechanisms to remind the person when that publication was posted in the first place, and prompt the end-user to reconsider sharing as outdated information could complicate response efforts \cite{wong2017social}. 

Although data permanence is a standard feature of many social media sites \cite{schlesinger2017situated}, researchers have argued about the virtues of ``forgetting'' as a design feature in social media applications \cite{xu2016automatic}. Providing ephemerality might be especially useful during disasters to eliminate information that is outdated and hence irrelevant.  It is important that social media platforms realize that in times of crisis cross-platform collaborations will happen; therefore, we believe that they should ensure to create cross-platform mechanisms to allow the flow of information and diminish the spread of outdated information that can potentially make response efforts more difficult.

\subsection{Educating People About Misinformation}
The third approach  was educating people about misinformation. Our investigation revealed how \textit{Verificado19S} and technologists  organized to create educational material  to educate people about how to self-verify information, as it informed the users of the best practices that they should follow before sharing something on social media.  Our survey also showed that participants were conscious about the importance of personally making sure that the information that they were about to share was accurate, and they took care to make photographic evidence  before sharing the information on multiple social computing platforms. This highlights how  people in \#19S were conscious about the importance of making sure the information was correct before spreading it on social media, and hence prevent misinformation. 
 
For future crises, social media platforms might consider revisiting this type of educational material and possibly use it to educate new populations affected by natural disasters. We believe it is also important to identify the differences this educational material has with traditional educational content. This might help identify an important gap in teaching people digital skills around misinformation.

\subsection{Trustworthy Organizations To Bridge  Government and Society}

An important difference that we found with previous work was the use of a  citizen-led initiative for coordinating relief efforts. Although \emph{Verificado19S} was very new, we believe it gained credibility and recognition because it worked with a collective of independent citizens from industry, nonprofits, and academia. It likely also helped that, as our macro-level analysis showcased, \emph{Verificado19S} operated independently from the government, which is generally distrusted in the Global South \cite{bargsted2017political}. We found similarities with a study in Ecuador \cite{wong2017social} in which Wong et al. observed an unwillingness of citizens to communicate or collaborate with the government to bring aid to affected areas. In our investigation, the patterns of communications across social media showed that they preferred to collaborate with \emph{Verificado19S}, in a citizen-driven approach for crisis response. The popularity of \emph{Verificado19S}, reflected by the news coverage it had \cite{HowtheVe71:online,Verifica17:online,AfterMex8:online,Verifica57:online,Abiertoa84:online}, and in our macro-analysis by the number of retweets, might have been precisely because it was an autonomous new organization that emerged organically in the \#19S aftermath. This highlights that during crisis relief work, we cannot always assume that citizens will be willing to share information with the government, as it has been found in other research work \cite{panagiotopoulos2016social,kaewkitipong2016community}.

However, we believe that due to  the social framework behind \emph{Verificado19S}, it is worth adopting a similar approach in other countries in which trustworthiness in governmental institutions is low, such as Latin America \cite{bargsted2017political}. 

\section{Design Implications}
Our macro-level analysis uncovered that ordinary citizens creating niche Facebook pages and groups that revolved around connecting people (e.g., to find missing individuals or their lost pets) were the ones with the most influence. Meanwhile, on Twitter, the most influential accounts belonged to organizations. These results highlight how the different affordances of social computing platforms might have facilitated certain types of self-organization for disaster response. Facebook's news feed algorithm promotes more personal content shared by ordinary people rather than news from organizations; while Twitter's algorithmic feed has put a greater focus on news and what others have liked or retweeted \cite{Nevermis91:online}. We believe it is important for practitioners and researchers to question: what interactions might be hindered by the algorithmic powers in play? and how might they affect the way people organize not only for disasters but for how our societies are constructed? Some social media platforms have already started to implement new affordances for times of crisis. For instance, Facebook allows citizens to mark they are safe during a crisis. However, we believe that platforms still have a lot of work to do to identify the type of affordances and social interactions that they want to facilitate during a crisis, especially the type of collaborations they want to facilitate to citizens across platforms. 

Regarding the spread of misinformation, we believe social media platforms should tackle the problem from a cross-platform perspective. Meaning that they should realize that in times of crisis, information will eventually flow from one platform to another. Therefore, it might be helpful during a crisis to incorporate ``watermarks'' onto the images and information that is shared online to help people contextualize when and where that data was generated. Our results showcased that people actively created their own ``watermarks'' to help others better identify when certain information came from another platform and might be now ``outdated''. We argue that social platforms should adopt similar approaches to mitigate the spread of inaccurate information across platforms. Having mechanisms to contextualize the information that people are exposed to, is especially important when thinking that people decide to conduct critical offline action based on this information. We invite social-technical platforms to consider having ``watermarks'' guidelines to implement during a crisis to help contextualize all information that is published during a crisis and help citizens maximize their time and energy in true concerns. 

Our micro-level analysis uncovered that people had ``backstage'' discussions on how to best present the \#19S news and needs that they planned to share. People wanted to share alarming news reports that might mobilize others to action. But they also did not want to create a sense of panic. As designers, it might help to consider tools that help people  to collectively ``frame'' how to present disaster news. 

Our paper also examined how more technical individuals contributed to earthquake relief efforts by collaborating on Slack and GitHub. These individuals created digital tools that seemed to bridge everyday users and technical ones. Our micro-level analysis revealed that they wanted to build a culture of prevention. They were interested in instituting a mentality in which people with  technical backgrounds were willing  to assist in rescue efforts during a crisis event. Some of these efforts have started to solidify within the country \cite{DatosAbi32:online}. Recently, the government launched an initiative to register volunteers that could help in the case of a crisis event, such as an earthquake. However, from our investigation, we argue that in order to be successful, these types of initiatives should come from independent civic response groups, as we found that citizens preferred to collaborate with an independent organization. This has been confirmed in previous work done in Ecuador \cite{wong2017social}. This work also found that individuals across Latin America have a lack of trust in the government, due to the history of corruption \cite{bargsted2017political}.

Our meso-level revealed that they reactivated their activities on Slack when there was an aftershock from \#19S. Future work could use these insights to design platforms that further encourage this collaboration and facilitate creating a community, which seemed important to technical folks. Previous research has found that people with previous connections are more effective during disasters \cite{messias2012latino}; social media platforms could adapt to allow people to connect with individuals with similar expertise and social circles to help connect during rescue efforts. 

We believe our findings could be useful to power tools that coordinate people within different social media platforms, not only for disaster response but for a range of endeavors, e.g., to counter misinformation in elections.

\section{Limitations}

This study confronted some methodological challenges that must be understood to interpret our findings correctly. It has been highlighted before \cite{hall2016following} that activities performed on different social media platforms are challenging to compare since they operate under different conceptual frameworks and motivations (e.g. retweeting on Twitter vs. mentioning a user on Slack). Therefore, the insights this work provides are constrained to these limitations. 

Our data analysis was constrained by the methodology used to collect the data. For example, the data from Slack was collected from the group behind the main website that was reported on in the media. While our data from Twitter was collected using the hashtags that the news reports mentioned people were using. The same is the case with our Facebook data, as it was collected from those groups that had in their name or description one or more of the \#19S related hashtags or keywords related to the earthquake. However, other hashtags could have been developed organically; if no media reported about them, it is likely that we missed gathering that data. Therefore, the quantitative analysis must be viewed from this perspective. 



For our interviews, we took care and made an effort to recruit participants from a range of social media sites. However, our pool of participants recruited from these sites likely belonged to particular cultural settings. We tried to counter this issue by triangulating their responses with quantitative social media data. Future work could focus on an analysis with a more varied population to overcome sample biases. Additionally, the social media platforms people used to contribute to earthquake relief methods in Mexico might not generalize to all other disasters. Future work could focus on studying how multiple platforms are utilized for different disasters across other latitudes. 

Notice that it is difficult to track how people jointly used multiple platforms, as people can use different usernames on each site \cite{hall2016following}. However, our quantitative and qualitative analysis allowed us to uncover first how specific individuals used multiple platforms jointly, in order to then better understand the patterns we visualized for each social computing platform. 

Finally, our understating of information verification process was limited by our methodology, future work could focus on conducting interviews to \textit{Verificado19S} participants to understand their workflow in greater detail.

\section{Conclusion}
In this paper, we examined social media use in the aftermath of the 7.1-magnitude earthquake that hit Mexico on September 19 of 2017. We analyzed the online interactions and collaborations that emerged across multiple platforms in the aftermath of the earthquake. We did a micro-analysis that allowed us to uncover the different purposes people had for using multiple platforms, which included building digital tools, mobilizing people offline and verifying news. We uncovered a citizen-driven approach to fact-check word of mouth reports about the situation in the city. We also uncovered how participants developed their own mechanisms to keep important information useful and understandable over time and across social networks. We also found that the groups in which people were posting the most amount of content and were sharing on social networks, such as Facebook, belonged  to categories in which people were requesting help. Therefore, having a way of sharing reliable information might have helped to streamline rescue efforts. From our findings, we conclude that misinformation should be tackled as a cross-platform problem. This is, the information should stay ``relevant'' and ``trustworthy'' while traveling from one social media platform to another.  
We also found similarities with work done in the Global South about the reluctance of citizens to coordinate with the government for rescue efforts, due to the lack of trust in the government; a problem that has been an issue in the overall region. Our research suggests how using independent, trustworthy organizations can be an alternative in these cases.  

\section{Acknowledgements}
Special thanks to Codeando M\'{e}xico, particularly to the former-director Miguel Salazar  for giving us access to the community.  Thanks to Andrés Monroy-Hern\'{a}ndez, Tonmona Tonny Roy, Juan Pablo Flores, Victor Romero, and No\'{e} Dominguez for their feedback on this work. We also would like to thank the anonymous reviewers who helped us improve the paper. This work was partially supported by NSF grant FW-HTF-19541, NSF CiTER grant,  and a Facebook Emerging Scholar fellowship.

\bibliographystyle{spmpsci}      

\bibliography{sample-base}
%
%


\end{document}